\documentclass[aps,prl,superscriptaddress,twocolumn,nopacs,amsmath,amssymb,letter]{revtex4-1}

\usepackage{color}
\usepackage{graphicx}
\usepackage{dcolumn}
\usepackage{bm}
\usepackage{xspace}
\usepackage{lineno}

\setcitestyle{numbers,square}

\begin{document}

\title{Tuneable electron-magnon coupling of ferromagnetic surface states in PdCoO$_2$}

\author{F.~Mazzola}
\email{mazzola@iom.cnr.it}
\affiliation {SUPA, School of Physics and Astronomy, University of St Andrews, St Andrews KY16 9SS, United Kingdom}

\author{C.-M.~Yim}
\affiliation {SUPA, School of Physics and Astronomy, University of St Andrews, St Andrews KY16 9SS, United Kingdom}

\author{V.~Sunko}
\affiliation {SUPA, School of Physics and Astronomy, University of St Andrews, St Andrews KY16 9SS, United Kingdom}
\affiliation {Max Planck Institute for Chemical Physics of Solids, N{\"o}thnitzer Stra{\ss}e 40, 01187 Dresden, Germany}

\author{S.~Khim}
\affiliation {Max Planck Institute for Chemical Physics of Solids, N{\"o}thnitzer Stra{\ss}e 40, 01187 Dresden, Germany}

\author{P.~Kushwaha}
\affiliation {CSIR-National Physical Laboratory, Dr K S Kishnana Marg, New Delhi-110012, India}
\affiliation {Max Planck Institute for Chemical Physics of Solids, N{\"o}thnitzer Stra{\ss}e 40, 01187 Dresden, Germany}

\author{O.~J.~Clark}
\author{L.~Bawden}
\affiliation {SUPA, School of Physics and Astronomy, University of St Andrews, St Andrews KY16 9SS, United Kingdom}

\author{I.~Markovi{\'c}}
\affiliation {SUPA, School of Physics and Astronomy, University of St Andrews, St Andrews KY16 9SS, United Kingdom}
\affiliation {Max Planck Institute for Chemical Physics of Solids, N{\"o}thnitzer Stra{\ss}e 40, 01187 Dresden, Germany}

\author{D.~Chakraborti}
\affiliation {SUPA, School of Physics and Astronomy, University of St Andrews, St Andrews KY16 9SS, United Kingdom}
\affiliation {Max Planck Institute for Chemical Physics of Solids, N{\"o}thnitzer Stra{\ss}e 40, 01187 Dresden, Germany}

\author{T.~K.~Kim}
\affiliation{Diamond Light Source, Harwell Campus, Didcot, OX11 0DE, United Kingdom}

\author{M.~Hoesch}
\affiliation{Diamond Light Source, Harwell Campus, Didcot, OX11 0DE, United Kingdom}
\affiliation{DESY Photon Science, Deutsches Elektronen-Synchrotron, Notkestra{\ss}e 85, 22607 Hamburg, Germany}

\author{A.~P.~Mackenzie}
\affiliation {Max Planck Institute for Chemical Physics of Solids, N{\"o}thnitzer Stra{\ss}e 40, 01187 Dresden, Germany}
\affiliation {SUPA, School of Physics and Astronomy, University of St Andrews, St Andrews KY16 9SS, United Kingdom}

\author{P.~Wahl}
\affiliation {SUPA, School of Physics and Astronomy, University of St Andrews, St Andrews KY16 9SS, United Kingdom}

\author{P.~D.~C.~King}
\email{pdk6@st-andrews.ac.uk}
\affiliation {SUPA, School of Physics and Astronomy, University of St Andrews, St Andrews KY16 9SS, United Kingdom}


\begin{abstract}
Controlling spin wave excitations in magnetic materials underpins the burgeoning field of magnonics. Yet, little is known about how magnons interact with the conduction electrons of itinerant magnets, or how this interplay can be controlled. Via a surface-sensitive spectroscopic approach, we demonstrate a strong and highly-tuneable electron-magnon coupling at the Pd-terminated surface of the delafossite oxide PdCoO$_2$, where a polar surface charge mediates a Stoner transition to itinerant surface ferromagnetism. We show how the coupling can be enhanced 7-fold with increasing surface disorder, and concomitant charge carrier doping, becoming sufficiently strong to drive the system into a polaronic regime, accompanied by a significant quasiparticle mass enhancement. Our study thus sheds new light on electron-magnon interactions in solid-state materials, and the ways in which these can be controlled. 
\end{abstract}

\date{\today}

\maketitle 

\section{Introduction}
Low-dimensional systems offer enormous potential for stabilising and controlling novel magnetic states and textures~\cite{Burch:2018, Huang:2020}. However, magnetic fluctuations are known to destabilise long-range order in two-dimensional (2D) systems (the famous Mermin-Wagner theorem \cite{Mermin:1966}). While magnetic anisotropy can allow long-range order to develop again~\cite{Onsager:1944, Gibertini:2019},  fluctuations can still be expected to play a crucial role. This necessitates their fundamental study, and potentially provides new opportunities in which to control magnonic excitations for spintronic technologies. 2D magnetic systems can be realised in a top-down approach, by exfoliating few/single layers from a bulk van der Waals magnet~\cite{Burch:2018,Gong:2017, Huang:2017}, or can be realised in a bottom-up approach, by realising magnetic order in a thin-film or surface geometry. The delafossite oxide, PdCoO$_2$, has recently been demonstrated as a model host of the latter~\cite{Mazzola:2018}. 

While it is a non-magnetic Pd $d^9$ metal in the bulk \cite{Mackenzie:2017}, the polar crystal structure (Fig.~\ref{f:overview}a) leads to pronounced charge carrier doping at the surface ~\cite{Sunko:2017,Mazzola:2018, Kim:2009, Noh:2009}. For the Pd-termination, the resulting self-doping is electron-like, which acts to shift a large peak in the unoccupied density of states towards the Fermi level. This in turn triggers a Stoner instability, generating a 2D ferromagnetic surface layer, as predicted by density-functional theory~\cite{Kim:2009,Mazzola:2018} and confirmed from electronic structure~\cite{Mazzola:2018} and anomalous Hall measurements~\cite{Harada:2019}. This thus provides a model environment, accessible to spectroscopic probes, in which to study the influence of magnetic excitations on the electronic structure of a 2D magnet. Here, we use angle-resolved photoemission (ARPES) and scanning-tunnelling microscopy and spectroscopy (STM/S) to investigate this system, finding evidence for a strong and highly-tuneable electron-magnon coupling.

\begin{figure*}[ht]
\includegraphics[width=\textwidth]{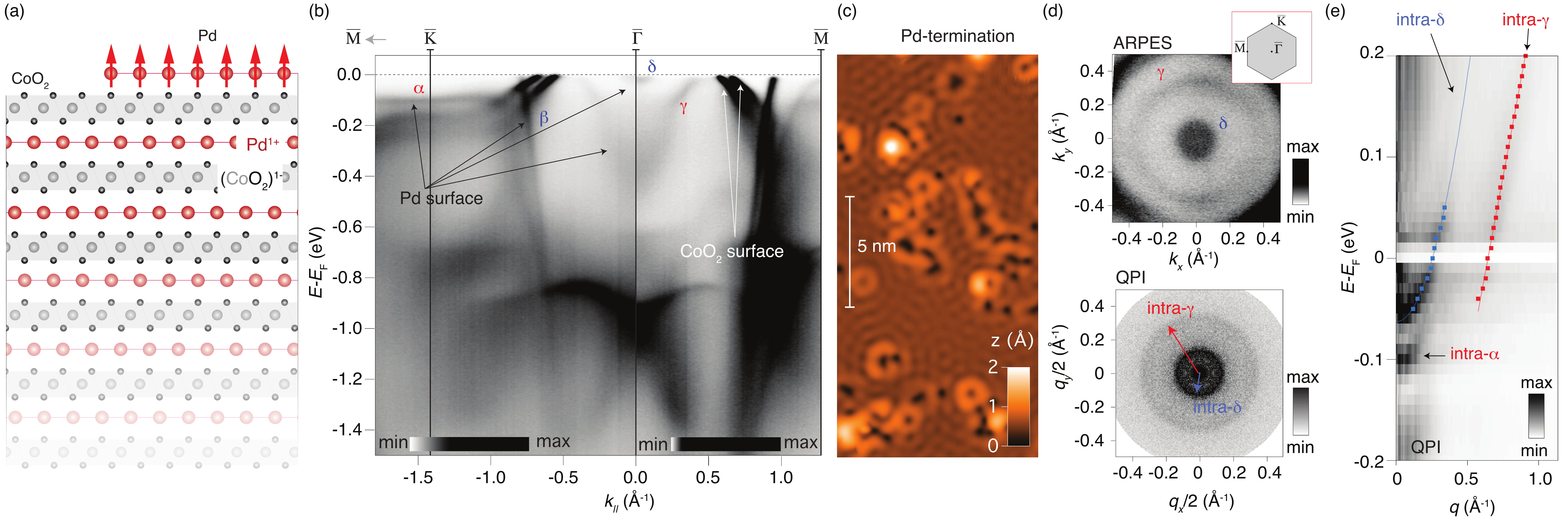}
\caption{{\bf Electronic structure of the Pd-terminated ferromagnetic surface of PdCoO$_2$.} (a) Schematic of the crystal structure (side view) of PdCoO$_2$. After sample cleaveage, a CoO$_2$ surface termination (left) and a Pd termination (right) are both present. (b) ARPES spectra ($h\nu=80$~eV along $\bar{\Gamma}$-\={M} and $h\nu=82$~eV along $\bar{\Gamma}$-\={K}) show the superposition of spectral weight arising from both terminations. The labelled $\alpha$-$\beta$ and $\gamma$-$\delta$ bands derive from the Pd-terminated surface, and represent exchange-split pairs by a surface ferromagnetism. (c) STM topographic image of a Pd-terminated region. (d) The STM quasiparticle interference map (bottom) from a Pd-terminated region is in good agreement with the Fermi surface measured by ARPES (top), determined by the $\gamma$ and $\delta$ bands. Signatures of such electron-like bands are also visible in energy-dependent QPI measurements (e) for intra-$\gamma$ and intra-$\delta$ band scattering, reported together with the corresponding fits. Additional spectral weight at 0.1~eV binding energy can be attributed to intra-$\alpha$ band scattering, in good agreement with the flat top of the $\alpha$ band observed by ARPES.}
\label{f:overview}
\end{figure*}

\section{Results}
\subsection{Electronic structure of the Pd-terminated ferromagnetic surface}
Figure~\ref{f:overview} shows an overview of the surface electronic structure of PdCoO$_2$ as measured by angle-resolved photoemission spectroscopy (ARPES) and scanning tunneling microscopy (STM). The crystal hosts two distinct surface terminations (Fig.~\ref{f:overview}a): a Pd-terminated surface and a CoO$_2$-terminated one. These would be expected with approximately equal probability, with a typical cleaved sample having both types present. Consistent with this, our measured ARPES spectra shown in Fig.~\ref{f:overview}b exhibit signatures of the electronic states associated with both surface terminations. A pair of heavy hole-like bands around the Brillouin zone centre derive from the CoO$_2$-terminated surface ~\cite{Sunko:2017, Noh:2009}, and will not be considered further here. Several additional low-energy states are observed, which we have previously attributed as deriving from the Pd-terminated surface~\cite{Mazzola:2018}. Of particular relevance here are the two pairs of exchange-split ferromagnetic surface states labelled ($\alpha$,$\beta$) and ($\gamma$,$\delta$) in Fig.~\ref{f:overview}b~\cite{Mazzola:2018}. The very flat band top of the spin-majority $\alpha$-band is visible $\approx100$~meV below $E_\mathrm{F}$. Its high associated density of states would sit at the Fermi level in the non-magnetic state; this is responsible for triggering a Stoner transition to itinerant surface ferromagnetism,~\cite{Mazzola:2018, Kim:2009} pushing this flat band below $E_\mathrm{F}$ as observed here.

To confirm that these key electronic states are indeed derived from the Pd-terminated surface, we have performed STM measurements to selectively probe a single surface termination (Fig.~\ref{f:overview}c) \cite{Yim:2021}. While the intrinsic defect density in the bulk is extremely low \cite{Sunko:2021}, we find a number of defects are present in our STM measurements. We tentatively attribute these to Pd vacancies at the cleaved surface ~\cite{Foot1}. Clear quasiparticle interference (QPI) patterns are visible around these, permitting a local measurement of the surface electronic structure. The Fourier transform of the QPI patterns, shown in Fig.~\ref{f:overview}d, exhibits two clear concentric circles, whose wavevector is in good agreement with intra-band scattering of the $\gamma$ and $\delta$ Fermi surfaces observed by ARPES. An energy-dependent cut through the QPI data (Fig.~\ref{f:overview}e) confirms this assignment from the electron-like character of these pockets, while an additional intense QPI signal at 100~meV below the Fermi level, peaked around $\mathbf{q}=0$ and with little evident dispersion, is consistent with the flat top of the $\alpha$ band observed in the ARPES. 

\subsection{Quasiparticle dynamics}
\begin{figure}
\includegraphics[width=\columnwidth]{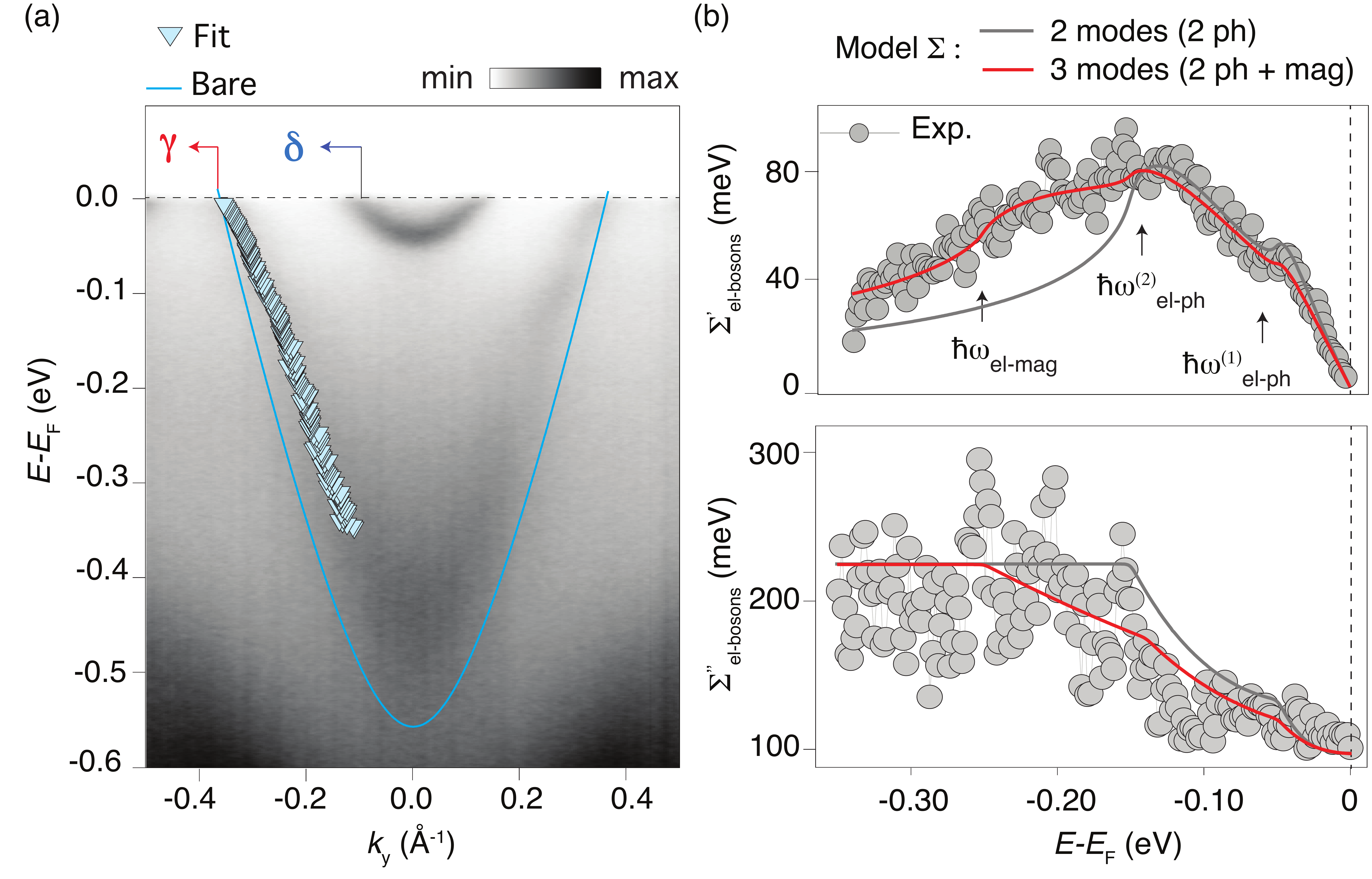}
\caption{{\bf Electron-boson coupling.} (a) Electronic structure measurements ($h\nu=90$~eV, measured along the $\bar{\Gamma}$-\={K} direction) showing the $\gamma$ and $\delta$ bands. The non-interacting, bare band, dispersion is taken as a $\mathbf{k}\cdot\mathbf{p}$ band, set to match the Fermi wavevectors of the experimental dispersions as described in the Methods. (b) The real and imaginary parts of the electron-boson self energy extracted from the data and from a Migdal-Eliashberg calculation as described in the Methods. A two-phonon model (grey lines) fails to describe the experimentally determined self energy, while a three-mode model (red line, with coupling to two phonons plus a magnon mode) is in much better agreement. The arrows in (b) show the characteristic mode energies. The real and imaginary parts of the self-energy retain causality through Kramers-Kronig transformation.}
\label{f:magnon}
\end{figure}

Having confirmed the electronic structure of the ferromagnetic Pd-terminated surface of PdCoO$_2$, we will focus now on spectroscopic signatures of marked electronic interactions evident for these surface states. Fig.~\ref{f:magnon}a shows in detail the $\gamma$ and $\delta$ bands. While the minority-spin $\delta$-band appears to be well described as a simple parabolic band, the $\gamma$-band exhibits several anomalies, or `kinks', in its dispersion. These are visible in the raw data, and in fits to momentum distribution curves (MDCs) shown as blue markers in Fig.~\ref{f:magnon}a. Typically, such spectral kinks are signatures of an electron-boson coupling, which can be described via an electronic self-energy. To investigate this further, we have extracted the self-energy from a self-consistent analysis of our ARPES measurements (see Methods), and report the Kramers-Kronig-consistent real and imaginary parts of the electron-boson self-energy in Fig.~\ref{f:magnon}b \cite{Foot2}

Two steep rises in the real-part of the electron-boson self-energy are visible at $\hbar\omega_{1}\approx50$~meV and at $\hbar\omega_{2}\approx130$~meV. These are in approximate agreement with phonon energies reported in the literature for PdCoO$_2$ \cite{Cheng:2017, Kumar:2013, Takatsu:2007}. The low-energy kinks of our measured data can thus be satisfactorily attributed to electron-phonon coupling. However, there is an additional broad hump in the extracted real part of the electron-boson self energy which extends out to $>300$~meV. This is an implausibly high energy to be due to a phonon mode (well above the highest phonon mode energies of $\approx 150$~meV expected for PdCoO$_2$ \cite{Cheng:2017, Kumar:2013, Takatsu:2007}). Indeed, a model self-energy calculation including only coupling to phonon modes fails to capture the high-energy part of the extracted experimental self energy for any electron-phonon coupling strength (the best fit is shown as the grey lines in Fig.~\ref{f:magnon}b; see also Supplementary Fig.~2). We thus conclude that there must be a third boson mode which is active here, with a characteristic energy scale of $>200$~meV, and which exhibits a marked coupling to the electronic sub-system. The high associated energy scale of this boson mode suggests that it may have a magnetic origin: in other itinerant ferromagnets, the spin-wave spectra are known to extend to similarly high energies of, for example, $>300$~meV in Fe, $>500$~meV in Co and $\approx500$~meV in Ni \cite{Halilov:1998, kubler:2009}.

\begin{figure*}
\includegraphics[width=0.8\textwidth]{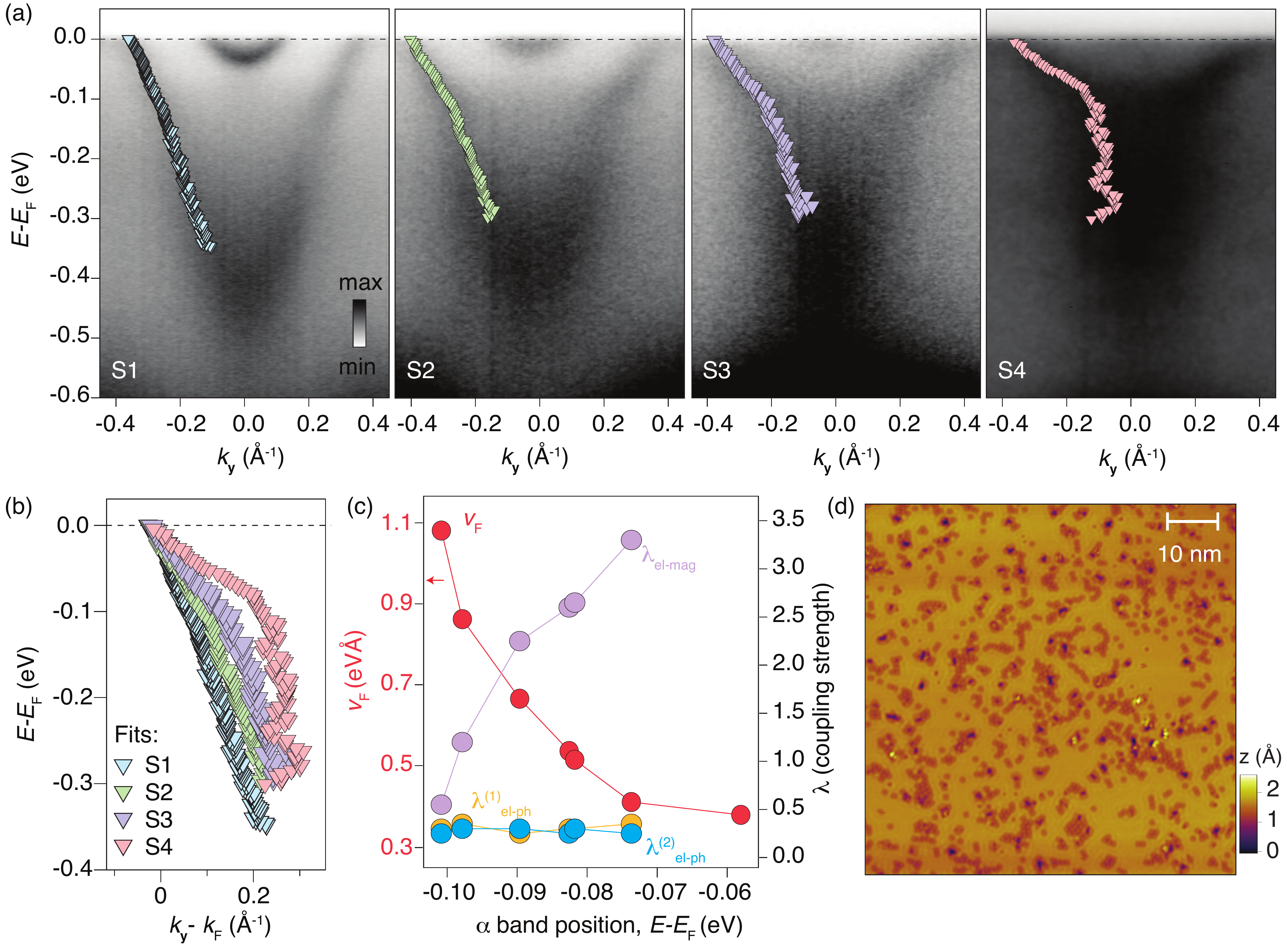}
\caption{{\bf Surface-dependent variation of electron-magnon coupling strengths.} (a) Electronic structure measurements of the $\gamma$ and $\delta$ bands as function of hole-doping. A pronounced increase in the band renormalisation is observed from the left to the right measurement, which were performed on different patches of the same sample (S1 and S2) and on different samples (S3 and S4). (b) Fits of MDCs (shown here as a function of $k_y-k_\mathrm{F}$; shifts in $k_\mathrm{F}$ of the $\gamma$ band between the samples are not resolvable experimentally)} show a strong increase in effective mass at the Fermi level (decrease in the Fermi velocity). By tracking the binding energy of the flat portion of the $\alpha$ band (see Supplementary Fig.~3), we find that this is correlated with an increasing hole doping (c). Fitting the extracted self energy for the these and additional samples, we find that the increased Fermi velocity (red) results due to a large increase in electron-magnon coupling strength, while the electron-phonon coupling strength does not vary significantly. (d) Large-scale topography measured by STM shows a large concentration of impurities (likely Pd vacancies) distributed across the cleaved Pd-terminated surface, which likely give rise to the variable hole doping determined above.
\label{f:variation}
\end{figure*}
Previous ARPES measurements have observed signatures of the coupling of electrons with such magnetic excitations in elemental magnetic metals~\cite{Hayashi:2013, Mynczak:2019, Cui:2007, Schneider:2016}, including for their two-dimensional surface electronic states~\cite{Hofmann:2009, Claessen:2004}, with spectroscopic signatures and energy scales very similar to those observed here. This can be understood via a process similar to the observation of electron-phonon coupling in ARPES: upon photoemission from a majority-state band, the resulting photohole can be filled by an electron with opposite spin accompanied by the emission or absorption of a magnon. Nonetheless, the different dispersion relations of phonons and magnons leads to differences in the functional form of the corresponding self-energy (see Methods). While we cannot describe our measured self-energies solely using electron-phonon-based models, extending them to include electron-magnon coupling leads to excellent agreement with our experimentally-determined self-energies (Fig.~\ref{f:magnon}b). The determined characteristic mode energy (left as a free parameter in fits to our experimental self-energies) is at 245~meV, while it exhibits a moderate coupling strength, $\lambda_{\mathrm{el-mag}}=0.55\pm0.05$, similar to, but slightly higher than, the coupling strength to the two phonon modes at lower energies:  $\lambda^{(1)}_{\mathrm{el-ph}}=0.30\pm0.02$ and $\lambda^{(2)}_{\mathrm{el-ph}}=0.25\pm0.03$. These values are in rather good agreement with calculated electron-paramagnon and electron-phonon coupling constants, respectively, in bulk elemental Pd \cite{Savrasov:1996, Bose:2008, Pinski:1979}, which is itself thought to be close to a ferromagnetic instability.    

Intriguingly, we find a strong sample-to-sample and spatial variation of the strength of the electron-magnon coupling. These are summarised in Fig.~\ref{f:variation}a (see also Supplementary Fig.~3). The electronic structure remains qualitatively the same as that shown in Fig.~\ref{f:overview}. However, the measured Fermi velocity extracted from fits to momentum distribution curves (Fig.~\ref{f:variation}b) decreases by a factor of $\approx3$ across these samples. At the same time, we observe a shift of the binding energy of the flat portion of the $\alpha$-band towards the Fermi level by $\approx{40}$~meV (Supplementary Fig.~4). The latter likely arises due to surface Pd vacancies. These are readily apparent in our STM measurements (Fig.~\ref{f:variation}d), and would act to partially counteract the polar surface charge, in turn leading to a reduction of the electron-like doping at the surface with respect to the hypothetical bulk-like surface termination. 

For the STM measurements presented in Fig.~\ref{f:variation}d, we estimate a surface Pd vacancy concentration of $\sim\!2\%$. This would lead to a nominal surface carrier density of 0.48 electrons/unit cell (el/u.c.), in good agreement with the Luttinger count of the ARPES-measured Fermi surfaces from sample S1 (Fig.~\ref{f:magnon}a). Assuming a rigid band shift with increasing Pd vacancy concentration, we estimate from tight-binding analysis (see Supplementary Fig.~5) that the variations observed across our samples here correspond to variations in the surface electron concentration of a further 0.02 el/u.c., corresponding to a further 2\% of vacancies in the surface Pd layer. While high concentrations of Pd vacancies would lead to exposed areas of CoO$_2$-terminated surface, the low concentrations of Pd vacancies observed here are thus best considered as isolated point defects, consistent with our STM measurements and providing a natural source of charge carrier doping.

\subsection{Tuneable electron-magnon coupling}
Taken together with the change in Fermi velocity, the above findings indicate a dramatic enhancement of the many-body renormalisations of the surface electronic structure with hole doping from surface Pd vacancies. From model fits to the electronic self energy from our ARPES measurements, we find that it is the magnon mode identified above which exhibits a strongly varying coupling strength, while the electron-phonon coupling remains unchanged within experimental error (Fig.~\ref{f:variation}c, see also Supplementary Fig.~6). The enhancement of this electron-magnon coupling is dramatic, reaching coupling strengths on the order of $\lambda\approx3$, where a weak-coupling Migdal-Eliashberg picture would be expected to break down.

\begin{figure}
\includegraphics[width=\columnwidth]{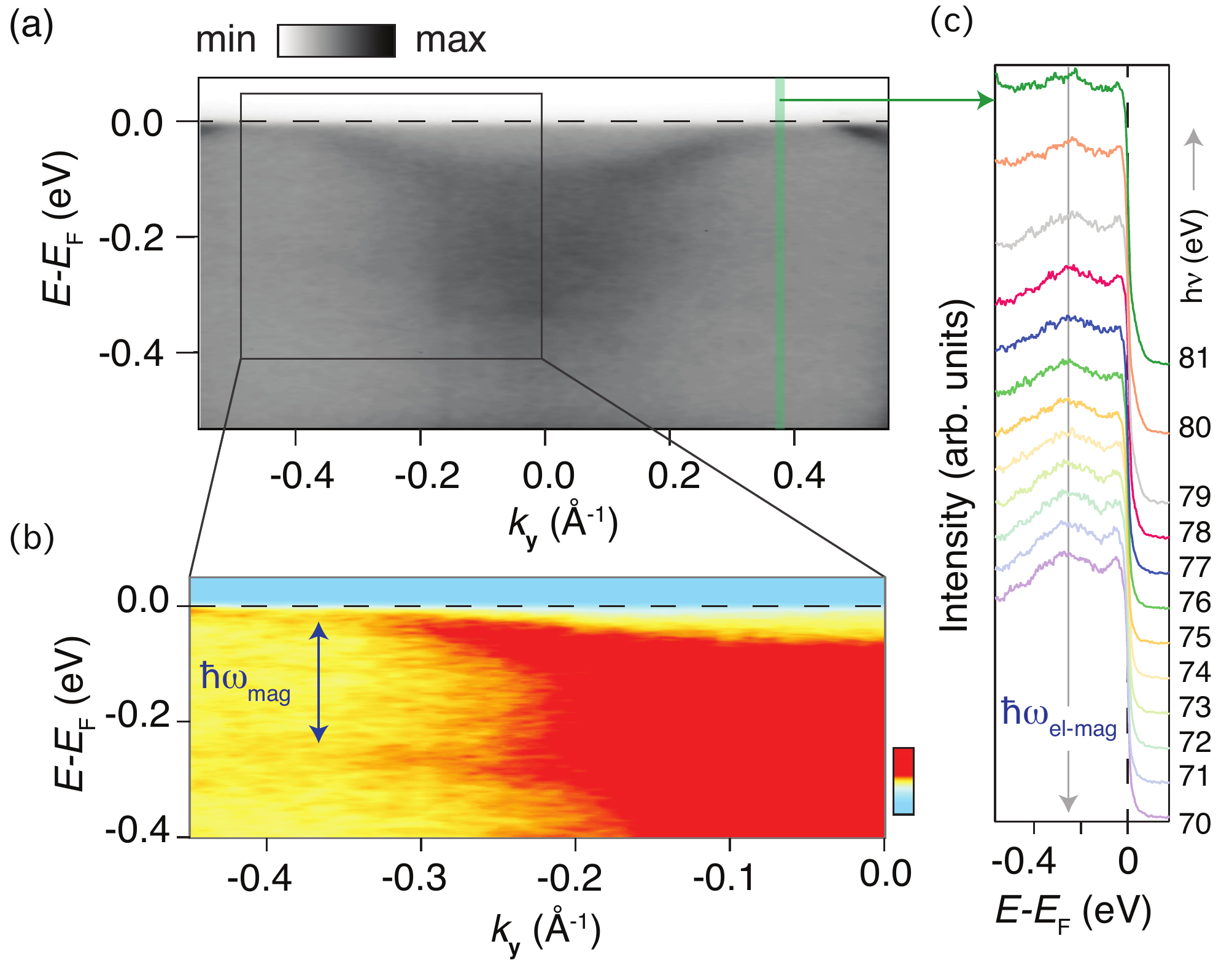}
\caption{{\bf Magnonic polarons.} (a) Measured $\gamma$-band dispersion for the sample with strongest electron-magnon coupling (S4, same data of Fig.~\ref{f:variation}a) but with incresed saturation and different color scale). The $\gamma$ band is now strongly renormalised, with the coherent quasiparticle band having an occupied bandwidth of only $\approx100$~meV (measurement taken at $h\nu=80$~eV). (b) A replica of this band is evident, shifted to higher binding energies by the characteristic magnon energy of $\approx240$~meV, indicative of polaron formation. (c) Energy distribution curves taken at the momentum value indicated by the green line in (a) show how this replica band feature is independent of photon energy, pointing to a strong intrinsic electron-magnon coupling regime.}
\label{f:pola}
\end{figure}
Indeed, for the sample marked S4 in Fig.~\ref{f:variation}, our weak-coupling models fail to adequately describe the extracted dispersion from our ARPES measurements. We show the measured dispersion again in Fig.~\ref{f:pola} with a different contrast. It is clear that the quasi-particle band is strongly renormalised, having an occupied band width of only $\approx100$~meV, significantly reduced compared to the $>400$~meV occupied band width for sample S1 seen in Fig.~\ref{f:magnon}. Within a Migdal-Eliashberg approach, this would imply a quasiparticle residue $Z=m_0/m^*<0.25$, a rather low value, and outside the regime of applicability of this approximation. Consistent with this, instead of simply generating a kink in the dispersion, we note that this strong electron-magnon coupling leads to a replica of the quasiparticle band, evident in our measurements as a weak dispersive feature with the same dispersion as the quasiparticle band, but shifted to higher binding energies by $\approx230$~meV (Fig.~\ref{f:pola}b). This is similar to the replica features observed due to strong coupling with phonons~\cite{Moser:2013, Lee:2014, Chen:2015, Wang:2016, Cancellieri:2016} or plasmons~\cite{Rliley:2018, Caruso:2021}, generating polaronic states. The separation between the quasiparticle band and the replica band observed here is equal to the characteristic mode energy of the magnon mode determined from fitting our data for lower coupling strengths (Fig.~\ref{f:variation}), and so we attribute this as a shake-off replica due to the strong electron-magnon coupling here. Its spectroscopic signatures are largely independent of photon energy (Fig.~\ref{f:pola}c). Together with the large quasiparticle mass renormalisation observed, this rules out that the replica band results from extrinsic photoelectron energy loss~\cite{Li:2018}, and instead indicates that the intrinsic electron-magnon coupling becomes strong enough to drive the system into a polaronic regime.

\section{Discussion}
Our measurements above indicate a dramatic enhancement of electron-magnon coupling due to increased disorder (vacancy concentration) of the cleaved surface. This can impact the coupling of magnons to the itinerant electrons in multiple ways. First, with the concomitant increased hole doping, the flat top of the $\alpha$ band is shifted towards the Fermi level (Fig.~\ref{f:variation}c). Correspondingly, the onset of the Stoner continuum from finite $\mathbf{q}$ spin-flip electron-hole excitations will be shifted to lower energies. The collective magnon mode will thus enter the Stoner continuum more rapidly, becoming Landau-damped and leading to a decreased quasiparticle lifetime, and thus enhanced coupling strength, as we observe here. We also note that the magnetism predicted in this surface layer by density-functional theory is only stabilised by a surface relaxation which increases the surface Pd-O bond length~\cite{Mazzola:2018}. The stability of the magnetic order, and as a consequence the strength of the electron-magnon coupling, is thus likely to be extremely sensitive to structural distortions and defects. Indeed, we find that the linewidth at the Fermi level also increases with hole doping here, a clear indicator of an enhanced electron-impurity scattering pointing to increased surface disorder which accompanies the increased electron-magnon coupling strength.

The detailed mechanisms by which the electron-magnon coupling becomes enhanced here require further detailed theoretical study. Nonetheless, our experimental results point to a surprisingly-large response of this coupling to small changes in carrier concentration and surface disorder. This firmly establishes the surface states of PdCoO$_2$ as a model system in which to investigate the coupling of itinerant electrons to spin excitations, of fundamental importance to understand the limits of stability of magnetic order in 2D. We hope that the findings presented here motivate future studies aimed at gaining true deterministic control over the large changes in electron magnon coupling which we have observed. In this respect, we note the enormous recent progress on the growth of thin films of PdCoO$_2$ \cite{Harada:2018, Ok:2020, Sun:2019, Brahlek:2019}, in some of which signatures of the ferromagnetism of the Pd-terminated surface have already been reported in transport measurements~\cite{Harada:2019}. Such thin films would be ideally suited for studies of gate-tuning of surface doping (or even surface disorder) via, {\it e.g.}, ionic liquid gating. Our results above show that for changes in the surface carrier concentration accessible to such gating techniques, the electron-magnon coupling at PdCoO$_2$ surfaces can be driven from a weak- to a strong-coupling polaronic regime, opening tantalising new possibilities for studying the creation and control of spin-polarons.

\section{Materials and Methods}
{\bf Angle-resolved photoemission:} Single-crystal samples of PdCoO$_2$ were grown via a flux method in sealed quartz tubes \cite{Tanaka:1996}. These were cleaved \textit{in situ} at the measurement temperature of $\approx10$~K. ARPES measurements were performed at the I05 beamline of Diamond Light Source, using a Scienta R4000 hemispherical electron analyser, and photon energies between 70 and 90~eV. All measurements used linear-horizontal ($p$-) polarised light. The lateral spot size of the photon beam on the sample is $\approx 50$~$\mu$m.

\

{\bf STM measurements:} The STM and tunneling spectroscopy experiments were performed using a home-built low temperature STM that operates at a base temperature down to 1.8 K \cite{White:2011}.  Pt/Ir tips were used, and conditioned by field emission on a gold target.   Differential conductance ($dI/dV$) maps and single point spectra were recorded by means of standard lock-in technique (see also Supplementary Fig.~7), with the frequency of the bias modulation set at 413 Hz.  The STM/S results reported here were obtained at a sample temperature of 4.2 K.

\

{\bf Self-energy analysis:} The electron-boson self-energy was extracted through a self-consistent analysis. The bare band is described by a nearly-parabolic $\mathbf{k} \cdot \mathbf{p}$ band, whose Fermi wavevectors are fixed to the experimental value. The parameters that describe the non-interacting band have been iteratively determined such that the causality connection between the extracted real and imaginary parts of the self energy enforced by Kramers-Kronig transformation is preserved. The electron-electron contribution to the self energy has been subtracted from the self-energies shown in Fig.~\ref{f:magnon}b to aid comparison of the electron-boson contribution. The full extracted self energy, including electron-electron contribution, is shown in Supplementary Fig.~1. 

The real and imaginary parts of the self-energy have been modeled using a Migdal-Eliashberg approach. The electron-phonon interaction was described by a conventional two-phonon Debye-model with characteristic cut-off energies. Finite temperature was also included in the model, as described in Ref. \cite{Mazzola:2013a, Mazzola:2017a}. For the electron-magnon coupling, the magnon dispersion relation, $\omega_{mgn}\propto q^2$, yields a density of states proportional to the square root of the energy, which means $D_{mgn}\propto \omega^{\frac{1}{2}}$. This manifests in a different shape of both real and imaginary parts of the self energy compared to the one expected for phonons (($D_{ph}\propto \omega$)). This approach is similar to what has been used in Ref. \cite{Hofmann:2009}. Under the Migdal-Eliashberg approach, the contribution of electron-boson coupling to the imaginary part of the self-energy is calculated as:

\begin{align} 
&\Im\Sigma^{electron-boson}(\omega,T)=\pi\int_{0}^{\omega_\mathrm{max}}\alpha^2(\omega^\prime)F(\omega^\prime)[1+2n(\omega^\prime) \nonumber \\
&\quad +\textit{f}(\omega+\omega^\prime)-f(\omega-\omega^\prime)] d{\omega^\prime} \nonumber
\end{align}
where $\omega_{\mathrm{max}}$ is the highest energy allowed for the boson mode, $f(\omega)$ and $n(\omega)$ are the fermion and boson distribution, respectively, and $T$ is the temperature. For the phonons, $\alpha^2 F(\omega)=0$ if $\omega>\omega_\mathrm{max}$  and $\alpha^2 F(\omega)=\lambda(\omega/\omega_\mathrm{max})^2$ for $\omega<\omega_\mathrm{max}$. In an analogous way, for magnons, $\alpha^2 F(\omega)=0$ if $\omega>\omega_\mathrm{max}$ and $\alpha^2 F(\omega)=\lambda_{mgn} (\omega/\omega_{mgn})^{\frac{1}{2}}$ for $\omega<\omega_\mathrm{max}$, where $\lambda_{mgn}$ represents the electron-magnon coupling strength of the system~\cite{Hofmann:2009}.

\section{Acknowledgements}
We thank C.~Hooley, T.~Frederiksen, G. van der Laan, G. Panaccione, H. Rosner and G. Siemann for useful discussions. We gratefully acknowledge support from the European Research Council (through the QUESTDO project, 714193), the Royal Society, the Max-Planck Society, and the UKRI Engineering and Physical Sciences Research Council (Grant No.~EP/S005005/1). We thank Diamond Light Source for access to Beamline I05 (Proposals SI12469, SI14927, and SI16262), which contributed to the results presented here. V.S., O.J.C., and L.B. acknowledge the EPSRC for PhD studentship support through Grants EP/L015110/1, EP/K503162/1, and EP/G03673X/1, respectively. I.M. and D.C. acknowledge studentship support from the International Max-Planck Research School for Chemistry and Physics of Quantum Materials.

\linespread{0.9}

\begin{thebibliography}{9}

\bibitem{Burch:2018}
Burch, K., Mandrus, D. \& Park, J. Magnetism in two-dimensional van der Waals materials. {\em Nature}. \textbf{563}, 47-52 (2018)

\bibitem{Huang:2020}
Huang, B., McGuire, M., May, A., Xiao, D., Jarillo-Herrero, P. \& Xu, X. Emergent phenomena and proximity effects in two-dimensional magnets and heterostructures. {\em Nature Materials}. \textbf{19}, 1276-1289 (2020)

\bibitem{Mermin:1966}

\bibitem{Onsager:1944}
Onsager, L. Crystal Statistics. I. A Two-Dimensional Model with an Order-Disorder Transition. {\em Phys. Rev.}. \textbf{65}, 117-149 (1944)

\bibitem{Gibertini:2019}
Gibertini, M., Koperski, M., Morpurgo, A. \& Novoselov, K. Magnetic 2D materials and heterostructures. {\em Nature Nanotechnology}. \textbf{14}, 408-419 (2019)

\bibitem{Gong:2017}
Gong, C., Li, L., Li, Z., Ji, H., Stern, A., Xia, Y., Cao, T., Bao, W., Wang, C., Wang, Y., Qiu, Z., Cava, R., Louie, S., Xia, J. \& Zhang, X. Discovery of intrinsic ferromagnetism in two-dimensional van der Waals crystals. {\em Nature}. \textbf{546}, 265-269 (2017)

\bibitem{Huang:2017}
Huang, B., Clark, G., Navarro-Moratalla, E., Klein, D., Cheng, R., Seyler, K., Zhong, D., Schmidgall, E., McGuire, M., Cobden, D., Yao, W., Xiao, D., Jarillo-Herrero, P. \& Xu, X. Layer-dependent ferromagnetism in a van der Waals crystal down to the monolayer limit. {\em Nature}. \textbf{546}, 270-273 (2017)

\bibitem{Mazzola:2018}
Mazzola, F., Sunko, V., Khim, S., Rosner, H., Kushwaha, P., Clark, O., Bawden, L., Marković, I., Kim, T., Hoesch, M., Mackenzie, A. \& King, P. Itinerant ferromagnetism of the Pd-terminated polar surface of PdCoO2. {\em Proceedings Of The National Academy Of Sciences}. \textbf{115}, 12956-12960 (2018)

\bibitem{Mackenzie:2017}
Mackenzie, A. The properties of ultrapure delafossite metals. {\em Reports On Progress In Physics}. \textbf{80}, 032501 (2017)

\bibitem{Sunko:2017}
Sunko, V., Rosner, H., Kushwaha, P., Khim, S., Mazzola, F., Bawden, L., Clark, O., Riley, J., Kasinathan, D., Haverkort, M., Kim, T., Hoesch, M., Fujii, J., Vobornik, I., Mackenzie, A. \& King, P. Maximal Rashba-like spin splitting via kinetic-energy-coupled inversion-symmetry breaking. {\em Nature}. \textbf{549}, 492-496 (2017)

\bibitem{Kim:2009}
Kim, K., Choi, H. \& Min, B. Fermi surface and surface electronic structure of delafossite PdCoO$_2$. {\em Phys. Rev. B}. \textbf{80}, 035116 (2009)

\bibitem{Noh:2009}
Noh, H., Jeong, J., Jeong, J., Cho, E., Kim, S., Kim, K., Min, B. \& Kim, H. Anisotropic Electric Conductivity of Delafossite PdCoO$_2$ Studied by Angle-Resolved Photoemission Spectroscopy. {\em Phys. Rev. Lett.}. \textbf{102}, 256404 (2009)

\bibitem{Harada:2019}
Harada, T., Sugawara, K., Fujiwara, K., Ito, S., Nojima, T., Takahashi, T., Sato, T. \& Tsukazaki, A. Anomalous Hall effect at the spontaneously electron-doped polar surface of PdCoO2 ultrathin films. {\em ArXiv:1908.08173}. (2019)

\bibitem{Yim:2021}
Yim, C., Chakraborti, D., Rhodes, L., Khim, S., Mackenzie, A. \& Wahl, P. Quasiparticle interference and quantum confinement in a correlated Rashba spin-split 2D electron liquid. {\em Science Advances}. \textbf{7} (2021)

\bibitem{Sunko:2021}
Sunko, V., McGuinness, P., Chang, C., Zhakina, E., Khim, S., Dreyer, C., Konczykowski, M., Borrmann, H., Moll, P., König, M., Muller, D. \& Mackenzie, A. Controlled Introduction of Defects to Delafossite Metals by Electron Irradiation. {\em Phys. Rev. X}. \textbf{10}, 021018 (2020)

\bibitem{Foot1}
Such surface defects could in principle also be due to residual H absorption~\cite{Torres:2015}, although we note that the STM measurements here were performed in cryogenic vacuum making that scenario less likely.

\bibitem{Foot2}
See Supplementary Fig.~1 for the full self energy with the electron-electron contribution included.



\bibitem{Cheng:2017}
Cheng, L., Yan, Q. \& Hu, M. The role of phonon–phonon and electron–phonon scattering in thermal transport in PdCoO2. {\em Phys. Chem. Chem. Phys.}. \textbf{19}, 21714-21721 (2017)

\bibitem{Kumar:2013}
Kumar, S., Gupta, H. \& Karandeep First principles study of structural, bonding and vibrational properties of PtCoO2, PdCoO2 and PdRhO2 metallic delafossites. {\em Journal Of Physics And Chemistry Of Solids}. \textbf{74}, 305-310 (2013)

\bibitem{Takatsu:2007}
Takatsu , H., Yonezawa , S., Mouri , S., Nakatsuji , S., Tanaka , K. \& Maeno , Y. Roles of High-Frequency Optical Phonons in the Physical Properties of the Conductive Delafossite PdCoO2. {\em Journal Of The Physical Society Of Japan}. \textbf{76}, 104701 (2007)

\bibitem{Halilov:1998}
Halilov, S., Eschrig, H., Perlov, A. \& Oppeneer, P. Adiabatic spin dynamics from spin-density-functional theory: Application to Fe, Co, and Ni. {\em Phys. Rev. B}. \textbf{58}, 293-302 (1998)

\bibitem{kubler:2009}
Kubler, J. Theory of Itinerant Electron Magnetism. {\em Oxford University Press}. (2009)

\bibitem{Hayashi:2013}
Hayashi, H., Shimada, K., Jiang, J., Iwasawa, H., Aiura, Y., Oguchi, T., Namatame, H. \& Taniguchi, M. High-resolution angle-resolved photoemission study of electronic structure and electron self-energy in palladium. {\em Phys. Rev. B}. \textbf{87}, 035140 (2013)

\bibitem{Mynczak:2019}
Młyńczak, E., Müller, M., Gospodarič, P., Heider, T., Aguilera, I., Bihlmayer, G., Gehlmann, M., Jugovac, M., Zamborlini, G., Tusche, C., Suga, S., Feyer, V., Plucinski, L., Friedrich, C., Blügel, S. \& Schneider, C. Kink far below the Fermi level reveals new electron-magnon scattering channel in Fe. {\em Nature Communications}. \textbf{10}, 505 (2019)

\bibitem{Cui:2007}
Cui, X., Shimada, K., Hoesch, M., Sakisaka, Y., Kato, H., Aiura, Y., Higashiguchi, M., Miura, Y., Namatame, H. \& Taniguchi, M. Angle-resolved photoemission spectroscopy study of Fe(110) single crystal: Many-body interactions between quasi-particles at the Fermi level. {\em Surface Science}. \textbf{601}, 4010-4012 (2007)

\bibitem{Schneider:2016}
Młyńczak, E., Eschbach, M., Borek, S., Minár, J., Braun, J., Aguilera, I., Bihlmayer, G., Döring, S., Gehlmann, M., Gospodari\ifmmode \checkc\else č\fi, P., Suga, S., Plucinski, L., Blügel, S., Ebert, H. \& Schneider, C. Fermi Surface Manipulation by External Magnetic Field Demonstrated for a Prototypical Ferromagnet. {\em Phys. Rev. X}. \textbf{6}, 041048 (2016)

\bibitem{Hofmann:2009}
Hofmann, A., Cui, X., Schäfer, J., Meyer, S., Höpfner, P., Blumenstein, C., Paul, M., Patthey, L., Rotenberg, E., Bünemann, J., Gebhard, F., Ohm, T., Weber, W. \& Claessen, R. Renormalization of Bulk Magnetic Electron States at High Binding Energies. {\em Phys. Rev. Lett.}. \textbf{102}, 187204 (2009)

\bibitem{Claessen:2004}
Schäfer, J., Schrupp, D., Rotenberg, E., Rossnagel, K., Koh, H., Blaha, P. \& Claessen, R. Electronic Quasiparticle Renormalization on the Spin Wave Energy Scale. {\em Phys. Rev. Lett.}. \textbf{92}, 097205 (2004)

\bibitem{Savrasov:1996}
Savrasov, S. \& Savrasov, D. Electron-phonon interactions and related physical properties of metals from linear-response theory. {\em Phys. Rev. B}. \textbf{54}, 16487-16501 (1996)

\bibitem{Bose:2008}
Bose, S. Electron–phonon coupling and spin fluctuations in 3d and 4d transition metals: implications for superconductivity and its pressure dependence. {\em Journal Of Physics: Condensed Matter}. \textbf{21}, 025602 (2008)

\bibitem{Pinski:1979}
Pinski, F. \& Butler, W. Calculated electron-phonon contributions to phonon linewidths and to the electronic mass enhancement in Pd. {\em Phys. Rev. B}. \textbf{19}, 6010-6015 (1979)

\bibitem{Moser:2013}Moser, S., Moreschini, L., Jaćć, J., Barišć, O., Berger, H., Magrez, A., Chang, Y., Kim, K., Bostwick, A., Rotenberg, E., Forro, L. \& Grioni, M. Tunable Polaronic Conduction in Anatase TiO$_2$. {\em Phys. Rev. Lett.}. \textbf{110}, 196403 (2013)

\bibitem{Lee:2014}
Lee, J., Schmitt, F., Moore, R., Johnston, S., Cui, Y., Li, W., Yi, M., Liu, Z., Hashimoto, M., Zhang, Y., Lu, D., Devereaux, T., Lee, D. \& Shen, Z. Interfacial mode coupling as the origin of the enhancement of Tc in FeSe films on SrTiO3. {\em Nature}. \textbf{515}, 245-248 (2014)

\bibitem{Chen:2015}
Chen, C., Avila, J., Frantzeskakis, E., Levy, A. \& Asensio, M. Observation of a two-dimensional liquid of Fröhlich polarons at the bare SrTiO3 surface. {\em Nature Communications}. \textbf{6}, 8585 (2015)

\bibitem{Wang:2016}
Wang, Z., McKeown Walker, S., Tamai, A., Wang, Y., Ristic, Z., Bruno, F., Torre, A., Riccò, S., Plumb, N., Shi, M., Hlawenka, P., Sánchez-Barriga, J., Varykhalov, A., Kim, T., Hoesch, M., King, P., Meevasana, W., Diebold, U., Mesot, J., Moritz, B., Devereaux, T., Radovic, M. \& Baumberger, F. Tailoring the nature and strength of electron–phonon interactions in the SrTiO3(001) 2D electron liquid. {\em Nature Materials}. \textbf{15}, 835-839 (2016)

\bibitem{Cancellieri:2016}
Cancellieri, C., Mishchenko, A., Aschauer, U., Filippetti, A., Faber, C., Barišić, O., Rogalev, V., Schmitt, T., Nagaosa, N. \& Strocov, V. Polaronic metal state at the LaAlO3/SrTiO3 interface. {\em Nature Communications}. \textbf{7}, 10386 (2016)

\bibitem{Rliley:2018}
Riley, J., Caruso, F., Verdi, C., Duffy, L., Watson, M., Bawden, L., Volckaert, K., Laan, G., Hesjedal, T., Hoesch, M., Giustino, F. \& King, P. Crossover from lattice to plasmonic polarons of a spin-polarised electron gas in ferromagnetic EuO. {\em Nature Communications}. \textbf{9}, 2305 (2018)

\bibitem{Caruso:2021}
Caruso, F., Amsalem, P., Ma, J., Aljarb, A., Schultz, T., Zacharias, M., Tung, V., Koch, N. \& Draxl, C. Two-dimensional plasmonic polarons in n-doped monolayer MoS$_2$. {\em Phys. Rev. B}. \textbf{103}, 205152 (2021)

\bibitem{Li:2018}
Li, F. \& Sawatzky, G. Electron Phonon Coupling versus Photoelectron Energy Loss at the Origin of Replica Bands in Photoemission of FeSe on SrTiO$_3$. {\em Phys. Rev. Lett.}. \textbf{120}, 237001 (2018)

\bibitem{Harada:2018}
Harada, T., Fujiwara, K. \& Tsukazaki, A. Highly conductive PdCoO2 ultrathin films for transparent electrodes. {\em APL Materials}. \textbf{6}, 046107 (2018)

\bibitem{Ok:2020}
Ok, J., Brahlek, M., Choi, W., Roccapriore, K., Chisholm, M., Kim, S., Sohn, C., Skoropata, E., Yoon, S., Kim, J., Lee, H. Pulsed-laser epitaxy of metallic delafossite PdCrO$_2$ films. {APL Materials}. \textbf{8}, 051104 (2020)

\bibitem{Sun:2019}
Sun, J., Barone, M., Chang, C., Holtz, M., Paik, H., Schubert, J., Muller, D., Schlom, D. Growth of PdCoO2 by ozone-assisted molecular-beam epitaxy. {\em APL Materials}. \textbf{7}, 121112 (2019)

\bibitem{Brahlek:2019}
Brahlek, M., Rimal, G., Ok, J., Mukherjee, D., Mazza, A., Lu, Q., Lee, H., Ward, T., Unocic, R., Eres, G. \& Oh, S. Growth of metallic delafossite PdCoO$_2$ by molecular beam epitaxy. {\em Phys. Rev. Materials}. \textbf{3}, 093401 (2019)

\bibitem{Tanaka:1996}
Tanaka , M., Hasegawa , M. \& Takei , H. Growth and Anisotropic Physical Properties of PdCoO 2 Single Crystals. {\em Journal Of The Physical Society Of Japan}. \textbf{65}, 3973-3977 (1996)

\bibitem{White:2011}
White, S., Singh, U. \& Wahl, P. A stiff scanning tunneling microscopy head for measurement at low temperatures and in high magnetic fields. {\em Review Of Scientific Instruments}. \textbf{82}, 113708 (2011)

\bibitem{Mazzola:2013a}
Mazzola, F., Wells, J., Yakimova, R., Ulstrup, S., Miwa, J., Balog, R., Bianchi, M., Leandersson, M., Adell, J., Hofmann, P. \& Balasubramanian, T. Kinks in the $\sigma$ Band of Graphene Induced by Electron-Phonon Coupling. {\em Phys. Rev. Lett.}. \textbf{111}, 216806 (2013)

\bibitem{Mazzola:2017a}
Mazzola, F., Frederiksen, T., Balasubramanian, T., Hofmann, P., Hellsing, B. \& Wells, J. Strong electron-phonon coupling in the $\sigma$ band of graphene. {Phys. Rev. B}. \textbf{95}, 075430 (2017)

\bibitem{Torres:2015}
Fernández-Torres, L., Sykes, E., Nanayakkara, S. \& Weiss, P. Dynamics and Spectroscopy of Hydrogen Atoms on Pd 111. {\em The Journal Of Physical Chemistry B}. \textbf{110}, 7380-7384 (2006)

\end{thebibliography}

\clearpage
\setcounter{figure}{0}
\makeatletter 
\renewcommand{\thefigure}{S\@arabic\c@figure}
\makeatother
\begin{figure*}
\includegraphics[width=0.5\textwidth]{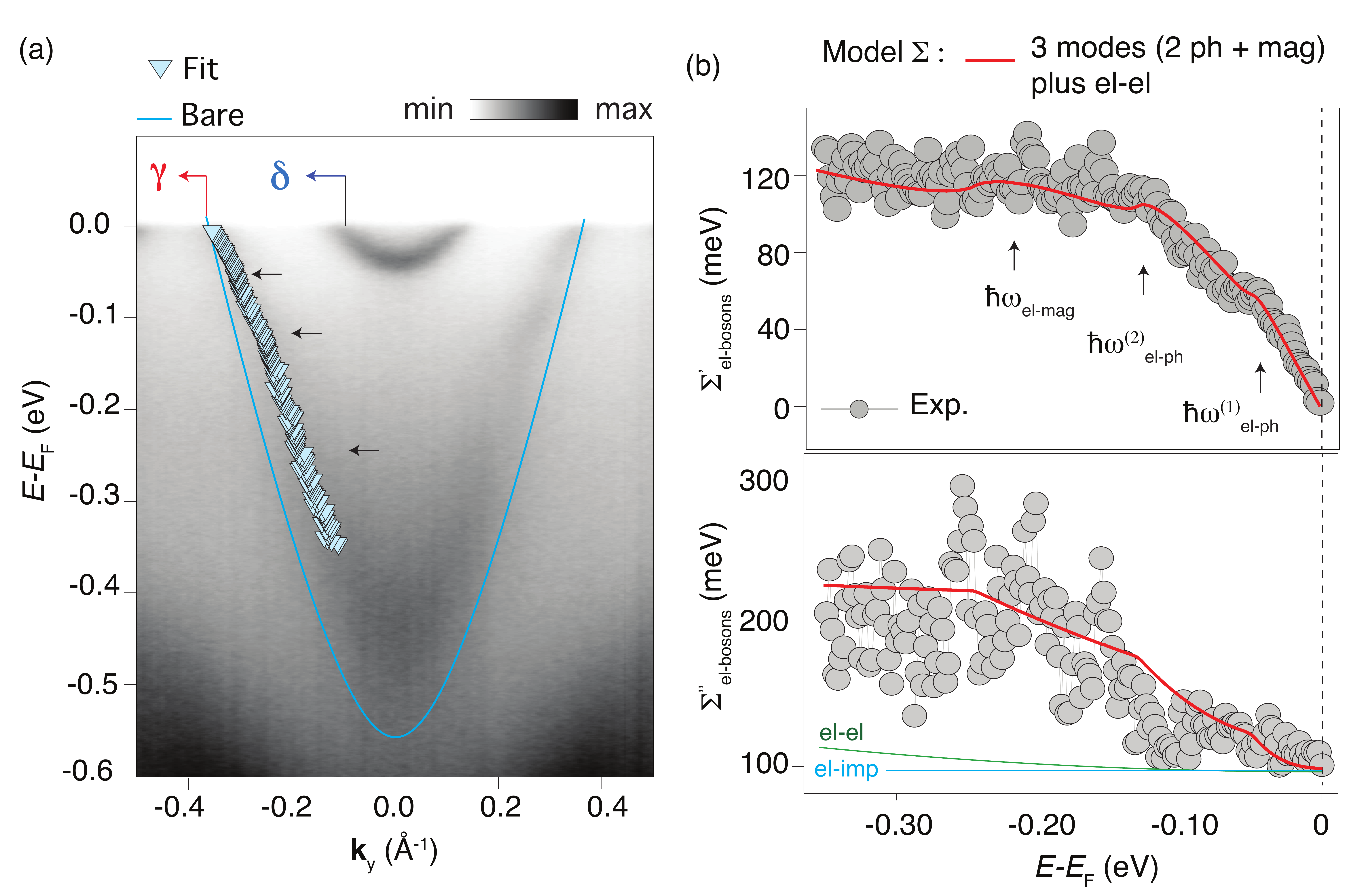}
\caption{{\bf Self-energy analysis.} (a) Electronic structure measurements at $90$~eV photon energy showing the $\gamma$ and $\delta$ bands, reproduced from Fig.~2 of the main text. (b) Full real and imaginary parts of the self energy extracted from the data. The three-mode model (red line) is also shown, including a magnon at $245$~meV and two phonons closer to the Fermi level, i.e. at $130$~meV and $50$~meV as described in the text. Electron-electron and electron-impurity scattering has also been included. The former manifests as an increase in the tail of both the real and imaginary parts of the self energy, as indicated in (b) by the green line. The real and imaginary parts retain causality through Kramers-Kronig transformations.}
\label{f:overview}
\end{figure*}

\begin{figure*}
\includegraphics[width=0.3\textwidth]{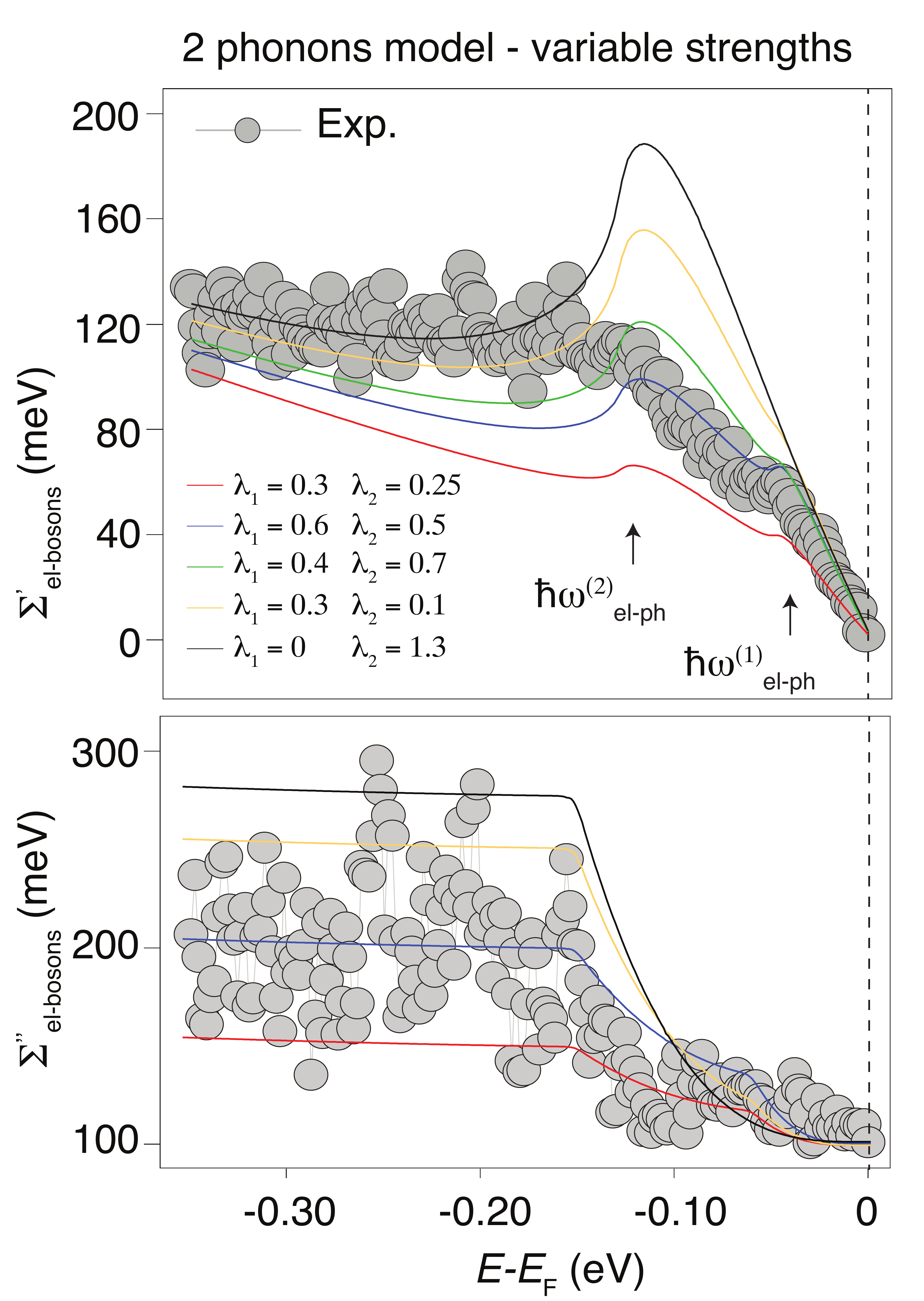}
\caption{{\bf Modelling including only phonon modes.} Real and imaginary parts of the electron-boson self energy extracted from the data (gray markers) along with simulated counterparts with inclusion of only 2 phonon modes, as described in the main text. The electron-phonon coupling strengths have been varied to attempt to obtain the best match to the experimental results, however even with arbitrary variation of the coupling strengths, it is not possible to simultaneously satisfactorily describe the real and imaginary parts of the self energy and their energy dependence. Similar lack of agreement was found with further variation of the mode energies, up to a maximum of $150$~meV which corresponds to the extreme maximum of the acoustic branch. Thus we conclude that with only phonon-based models, we cannot describe our experimental observations.}
\label{f:overview}
\end{figure*}

\begin{figure*}
\includegraphics[width=0.9\textwidth]{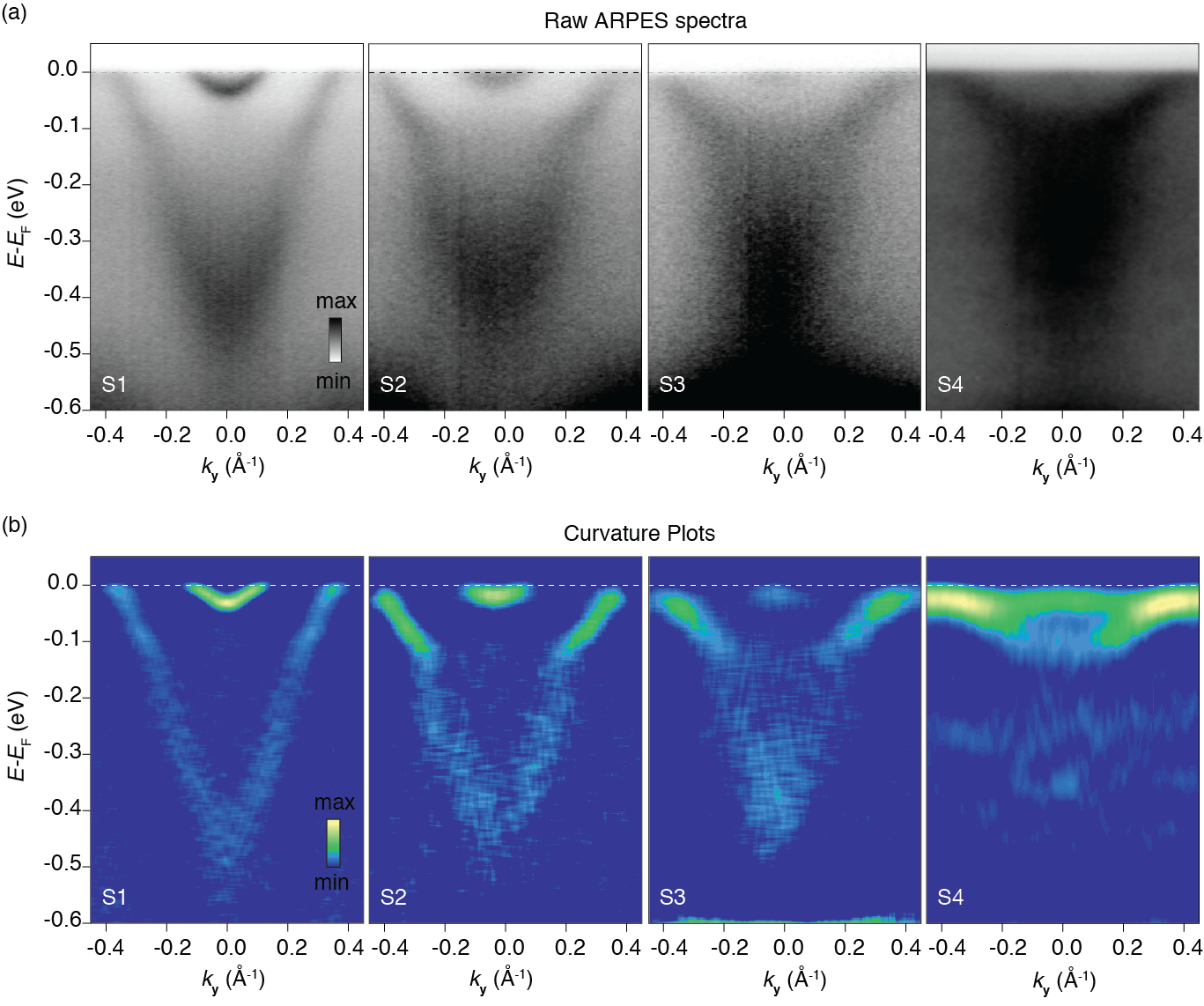}
\caption{{\bf Curvature analysis of ARPES data}. (a) ARPES measurements reproduced from Fig. 3(a) of the main text and (b) corresponding curvature analysis\footnote{Zhang, P. {\it et al.}, A precise method for visualizing dispersive features in image plots. Rev. Sci. Instrum. 82, 043712 (2011).} of the measured dispersions. The trends visible in the curvature data support the conclusions from the quantitative analysis presented in the main text, including: the multi-kink structure of sample S1; the increase in coupling strength from samples S1 to S4 leading to a pronounced decrease in Fermi velocity; the disappearance of the $\delta$-pocket with doping; and the emergence of a replica band feature split off from the main quasiparticle peak by ~230 meV in sample S4. }
\label{f:overview}
\end{figure*}

\begin{figure*}
\includegraphics[width=0.5\textwidth]{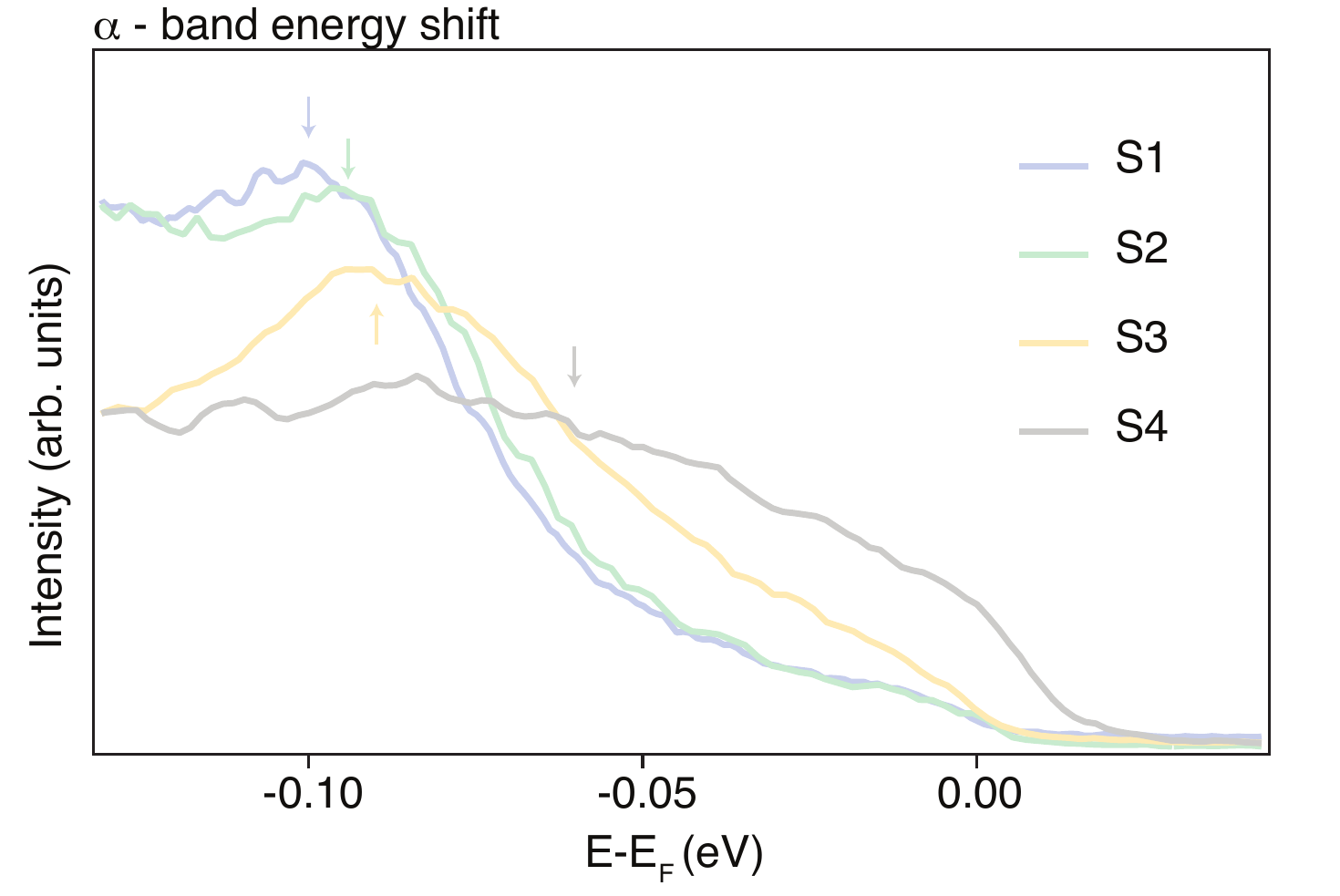}
\caption{{\bf Sample-dependent doping variations.} Energy distribution curves from samples S1-S4 shown in the main text (Fig. 3), taken at a momentum $-1.35\pm0.015$\AA$^{-1}$, cutting through the flat portion of the $\alpha$ band. A clear shift in the onset of this band towards the Fermi level is observed (indicated by the arrows), consistent with the p-type doping effect discussed in the main text. An increased linewidth is also visible, consistent with an increase in electron-impurity scattering rate due to surface disorder.}
\label{f:overview}
\end{figure*}
\begin{figure*}
\includegraphics[width=0.9\textwidth]{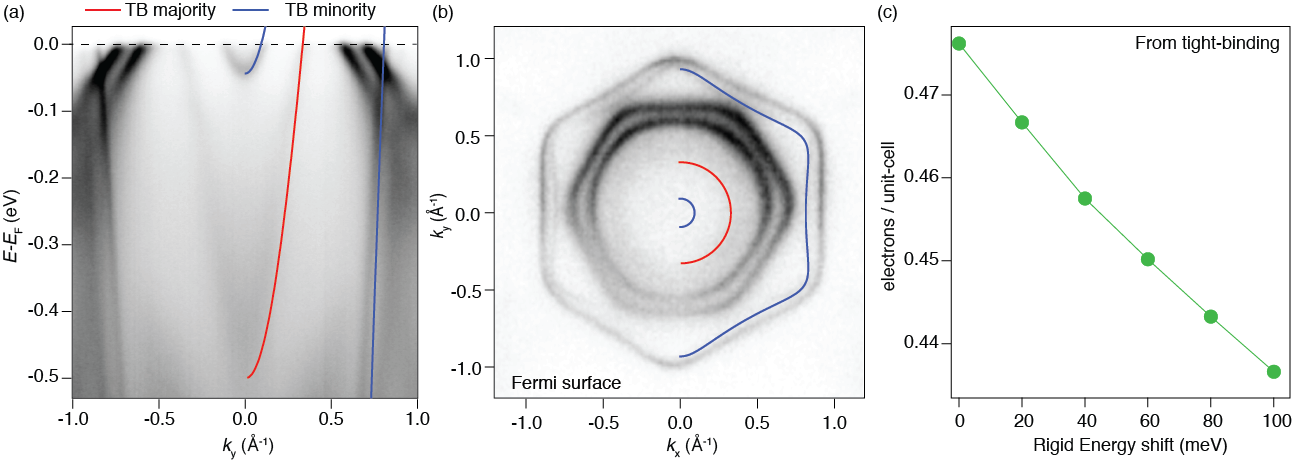}
\caption{{\bf Tight Binding Analysis}. (a,b) A simple tight-binding parametrization of the Pd-derived surface states which cross the Fermi level, shown superimposed on the ARPES data from sample S1 of the main text. Blue and red colors indicate the majority and minority spin components, respectively. Our tight-binding model agrees well with the ARPES data both for (a) energy-momentum dispersion and (b) the surface Fermi surface. (c) Corresponding change in surface carrier density for these tight-binding bands as a function of a shift of the chemical potential. Shifts of the chemical potential on the order of $40$~meV, as observed experimentally between samples S1 to S4, correspond to doping changes on the order of $0.02$~electrons/unit cell, which would arise from a change in surface Pd vacancy concentration of ~2\% between our measured samples; an entirely reasonable value to expect from cleave-to-cleave variations. We note that this analysis neglects the changes in quasi-particle velocity due to the varying many-body interactions which we observe, as well as any potential feedback of the doping on the resulting exchange splitting of the electronic states in the system. However, this should make only small changes in the calculated doping vs. band energy shift, which yields values consistent with direct estimation of carrier densities from the measured ARPES data.}
\label{f:overview}
\end{figure*}

\begin{figure*}
\includegraphics[width=0.7\textwidth]{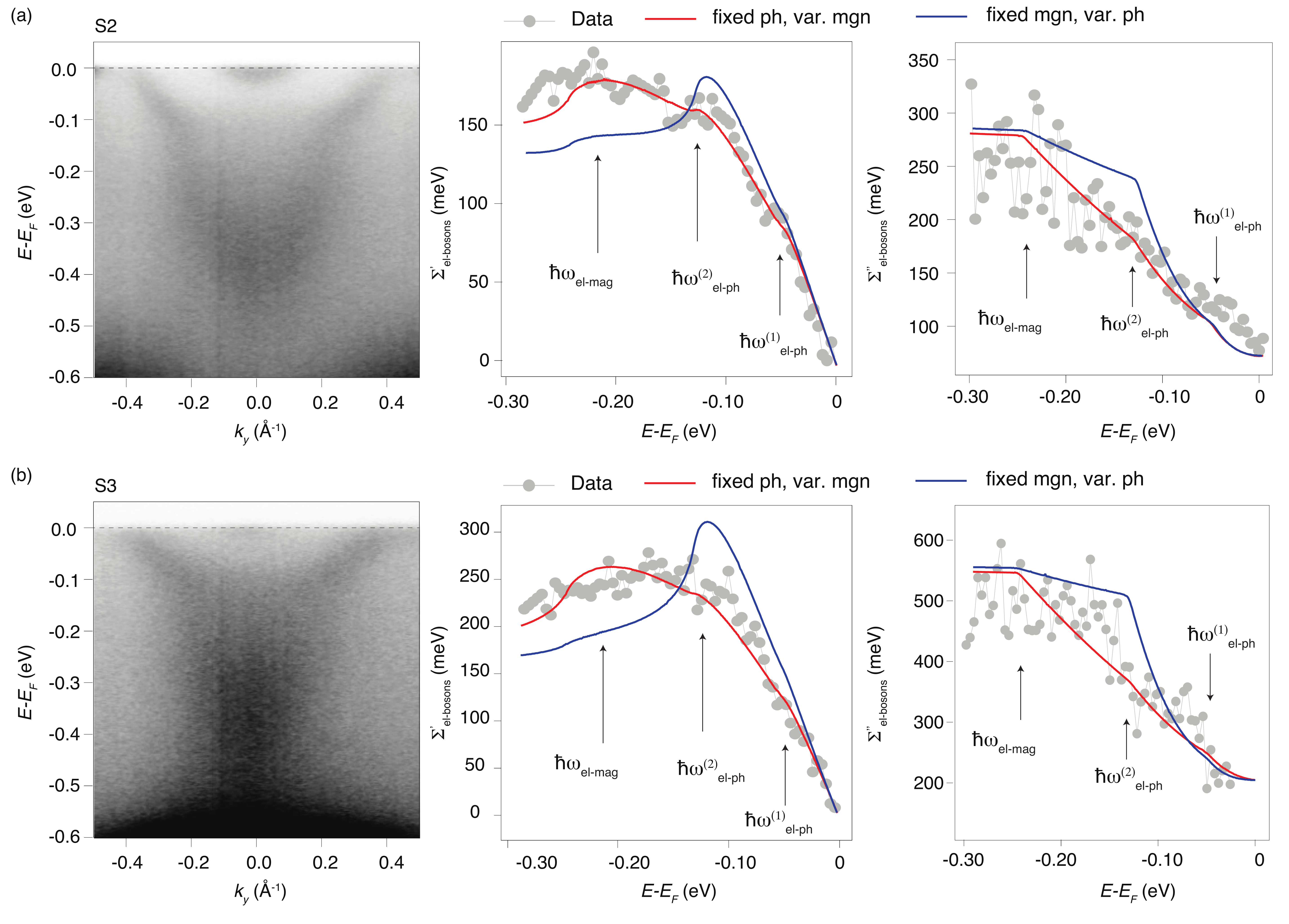}
\caption{{\bf Doping-dependent self-energy variations.} Self energy extraction for (a) S2 and (b) S3 samples of the main text. The self energy components have been fitted in two ways: The first, indicated by the red model, is obtained by allowing the strength of all the electron-boson couplings to vary. This model, regardless of the samples, gives a very good description of the experimental data and a negligible variation of the electron-phonon coupling strength. The second fit (blue lines) is obtained by fixing the magnon strength to match the one extracted for S1 and allowing the electron-phonon coupling strength to vary. This model, irrespective of the strength of the electron-phonon coupling does not show a satisfactory model to explain the experimental points. We thus conclude that the main effects underpinning the changes observed experimentally are due to changes in the electron-magnon coupling strength, as described in the main text. }
\label{f:overview}
\end{figure*}

\begin{figure*}
\includegraphics[width=0.5\textwidth]{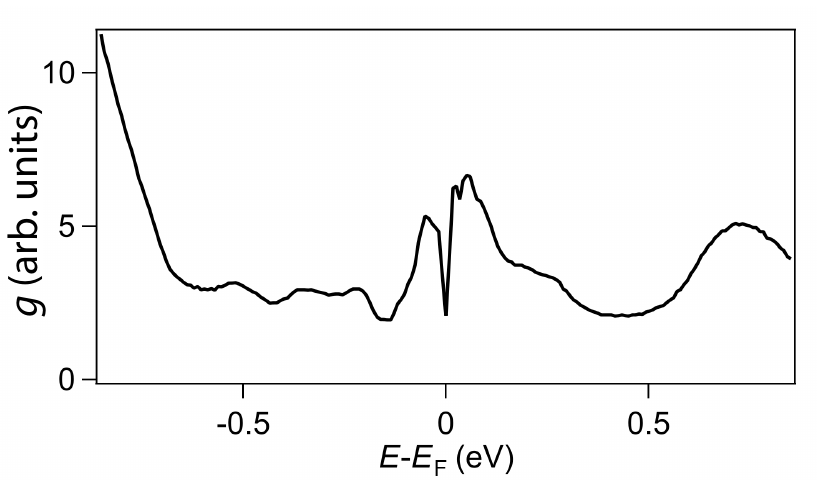}
\caption{{\bf Differential conductance.} Point differential conductance $dI/dV$ spectrum recorded from a defect-free position on the Pd-terminated surface, measured using the following parameters: spectroscopy set-point ($V_{SP}$, $I_{SP}$): 850~mV, 520~pA; amplitude and frequency of bias modulation: 6~mV, 413~Hz.}
\label{f:overview}
\end{figure*}

\end{document}